# Quantum frequency conversion and single-photon detection with lithium niobate nanophotonic chips

Xina Wang,[1,2,3] Xufeng Jiao,[1,2,3] Bin Wang,[1,2,3] Yang Liu,[2,3] Xiu-Ping Xie,[2,3] Ming-Yang Zheng,[2,3] Qiang Zhang,[1,2,3,*] and Jian-Wei Pan,[1,3]

[1]*Hefei National Research Center for Physical Sciences at the Microscale and School of Physical Sciences, University of Science and Technology of China, Hefei 230026, China*

[2]*Jinan Institute of Quantum Technology and CAS Center for Excellence in Quantum Information and Quantum Physics, University of Science and Technology of China, Jinan 250101, China*

[3]*Hefei National Laboratory, University of Science and Technology of China, Hefei 230026, China*

*qiangzh@ustc.edu.cn

**ABSTRACT:** In the past few years, the lithium niobate on insulator (LNOI) platform has revolutionized lithium niobate materials, and a series of quantum photonic chips based on LNOI have shown unprecedented performances. Quantum frequency conversion (QFC) photonic chips, which enable quantum state preservation during frequency tuning, are crucial in quantum technology. In this work, we demonstrate a low-noise QFC process on an LNOI nanophotonic platform designed to connect telecom and near-visible bands with sum-frequency generation by long-wavelength pumping. An internal conversion efficiency of 73% and an on-chip noise count rate of 900 counts per second (cps) are achieved. Moreover, the on-chip preservation of quantum statistical properties is verified, showing that the QFC chip is promising for extensive applications of LNOI integrated circuits in quantum information. Based on the QFC chip, we construct an upconversion single-photon detector with the sum-frequency output spectrally filtered and detected by a silicon single-photon avalanche photodiode, demonstrating the feasibility of an upconversion single-photon detector on-chip with a detection efficiency of 8.7% and a noise count rate of 300 cps. The realization of a low-noise QFC device paves the way for practical chip-scale QFC-based quantum systems in heterogeneous configurations.

## 1. INTRODUCTION

Photonic integrated chips with multiple optical operations have been extensively studied in the last decade for quantum information [1-4]. Among semiconductor-on-insulator platforms, including silicon [5], GaAs [6,7], lithium niobate [8,9], and others [10-12], LNOI is considered a promising integrated platform owing to its wide transparency window (350-4500 nm), outstanding electro-optic property, large optical nonlinearity efficiency, flexible ferroelectric domain control, and feasible heterogeneous integration with silicon [13-15], NbN [16] and $Si_3N_4$ [17] and so on [18]. As one of the building blocks for a quantum photonic circuit based on LNOI, the frequency converter has been investigated for its extremely high conversion efficiency owing to its strong optical confinement and state-of-the-art fabrication technology [19-21]. In chip-scale quantum information applications, a low-noise quantum frequency converter on LNOI is of the essence while unexplored thus far.

QFC has found extensive applications in quantum information research, including single-photon detectors [22-24], interfaces between long-distance quantum memories and various quantum systems [25-29], and photon-color erasers [30,31]. In these previous demonstrations, efficient quantum frequency converters were implemented with periodically poled lithium niobate (PPLN) waveguides with optical mode dimensions of several microns using ridge waveguides [32] or diffused waveguides fabricated by proton exchange [26,33] or titanium indiffusion [34]. The noise distribution of QFC pumped with telecom band laser has been studied on periodically poled LNOI nanophotonic chip [35]. The employed sum-frequency generation (SFG) process induced an insufferable noise count rate (NCR) at the

Mcps level mainly due to remarkable spontaneous Raman scattering (SRS).

In this work, we design and fabricate a low-noise QFC nanophotonic waveguide on an LNOI chip. The telecom band single photons are upconverted to the near-visible band by pumping with a 1950 nm single-frequency laser. An internal conversion efficiency of 73% and an on-chip NCR of 900 cps are achieved. Comparing the second-order correlation properties of the telecom-band heralded single photon source (HSPS) before and after the QFC, the quantum state preservation on the LNOI platform is verified. Such a QFC chip is promising for the extensive applications of LNOI integrated circuits in quantum information. With the sum-frequency output spectrally filtered and detected by a silicon single-photon avalanche photodiode (SPAD), we construct an upconversion single-photon detector (SPD) showing an average detection efficiency (DE) of 8.7% and an NCR of 300 cps, which demonstrates the feasibility of an LNOI-chip-based upconversion SPD.

## 2. CHIP FABRICATION AND CHARACTERIZATION

In this part, the QFC chip is designed and fabricated to realize frequency upconversion between the 1550 nm band (signal) and the 860 nm band (sum-frequency) with a single frequency pump laser fixed at 1950 nm as shown in Fig. 1(a). Then, the on-chip performances, including SFG efficiency and nonlinear noise, are demonstrated and analyzed. To verify the quantum state preservation, the second-order correlation properties of the HSPS are measured.

### A. Design and Fabrication

The device is fabricated on a Z-cut lithium niobate thin film bonded to a 2-μm thick thermally grown silicon dioxide layer on a silicon substrate (NANOLN Inc.). To utilize the maximum nonlinear coefficient of lithium niobate, which is $d_{33}\sim27$ pm/V, both the signal light and the pump light are expected to propagate in the TM mode along the X-axis. In Z-cut LNOI, two kinds of undesirable conditions can easily occur if the waveguide parameters are designed improperly, especially in our device where the three mixing lasers cover a wide wavelength span. The first is lateral leakage, attributed to the effective refractive index of the mode in the rib waveguide being lower than that in the slab, leading to the radiation of the guided mode in the shallow rib waveguide into orthogonally polarized slab modes [36], resulting in high propagation loss of the signal. As an example, the simulated mode field distribution is shown in Fig. 1(b) with an LN thickness of 800 nm, ridge height of 351.5 nm, and top width of 1.8 μm at 1550 nm. Generally, it can be prevented by increasing the ridge height for an LNOI film of a given thickness.

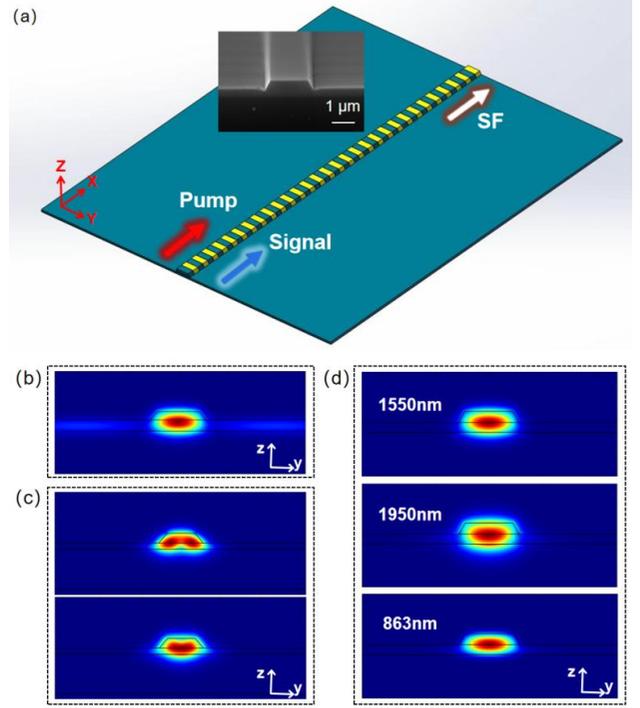

Fig. 1. (a) Schematic of the QFC chip. Inset is the scanning electron microscopy image of the fabricated waveguide. (b) Simulated profiles for the lateral leakage mode at 1550 nm. (c) Simulated profiles for hybrid modes at 1950 nm. (d) Simulated profiles for the quasi-TM mode of the signal, pump, and SF. The LN thickness, top width, ridge height, and sidewall angle are 800 nm, 1.8 μm, 422 nm, and 60°, respectively.

The second is the existence of special modes called "hybrid modes" caused by asymmetry of the structural refractive index distribution, which is closely related to waveguide parameters [37]. As an example, the profiles of two adjacent modes at 1950 nm are shown in Fig. 1(c), both of which are hybrid modes with an LN thickness of 800 nm, ridge height of 450 nm, and top width of 1.5 μm. Among these pump laser modes, the nondominant field components take a non-negligible proportion, and their power even equals the dominant field, which will greatly reduce the overlap integral of the three interacting waves. The simulation results show that when the ridge height exceeds a certain depth and gradually increases toward the film thickness, the TE polarization fraction defined by $\frac{\iint |E_y|^2 dydz}{\iint (|E_z|^2+|E_y|^2) dydz}$ of the pump quasi-TM mode tends to increase, while that of the sum-frequency (SF) quasi-TM mode tends to decrease. This unfavorable trade-off between the short wave and long wave limits the mode-overlap integral for three-wave interactions, which is closely related to the normalized nonlinear efficiency.

The effects of lateral leakage and hybrid modes can be greatly reduced by carefully selecting the waveguide parameters. Finally, we choose an LN thickness of 800 nm, a ridge height of 422 nm, and a top width of 1.8 μm, and correspondingly, the quasi-phase matching period is 4.1 μm. The simulated theoretical normalized conversion efficiency reaches 3552%/(W·cm$^2$) due to the optimized large mode-overlap integral and tight light confinement, although there is still a small amount of TE components in the quasi-TM modes for the SF and pump. The modes of the signal, pump, and SF are shown in Fig. 1(d).

To fabricate the nanophotonic chip, we first pattern the poling electrodes on the surface of the lithium niobate thin film using UV lithography and wet etching. Then, several 800 V pulses with a duration of tens of seconds are applied to achieve periodic domain inversion with a duty cycle of ~50%. After periodic poling, the electrodes are removed, and the waveguide is patterned by electron-beam lithography (EBL) following an Ar-based dry etching process to form a ridge waveguide [38]. An optimized RCA cleaning process is applied to remove the redeposition created in the dry etching process and reduce the selective etching to periodical domains. The scanning electron microscopy image of the fabricated waveguide is shown in the inset of Fig. 1(a). The ripples on the surface of the slab are caused by selective etching, while the side wall and upper surface of the waveguide are protected well. The length of the waveguide is 5.3 mm, and both facets are polished by chemical mechanical polishing.

## B. Frequency Conversion Performance

The measurement setup is depicted in Fig. 2(a). A homemade 1950 nm single-frequency laser and the signal light from a continuous-wave telecom tunable laser source (Santec TSL-550, 1480-1630 nm) are coupled into an in-line fiber polarization controller to ensure TM polarization separately. Then, the two beams are combined through a 1550-nm/1950-nm wavelength division multiplexer (WDM) and injected into the nano-waveguide with a lensed fiber (OZ optics). The mode field diameter of the lensed fiber is 2.5±0.5 μm, and the coupling loss is estimated to be ~4 dB for the signal wavelength by dual-edged coupling between the waveguide and two lensed fibers. The output is collected by an aspheric lens and collimated into free space.

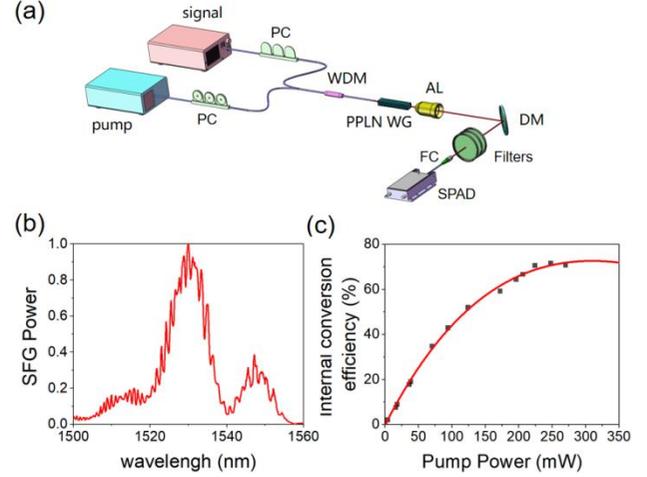

Fig. 2. (a) Experimental setup. PC, polarization controller; WDM, wavelength division multiplexer; WG, waveguide; AL, aspherical lens; DM, dichroic mirror; Filters, including a longpass filter, a shortpass filter, two bandpass filters, and an etalon; FC, fiber coupler; SPAD, silicon single-photon avalanche photodiode. (b) SFG tuning curve. (c) measured SFG conversion efficiency versus pump power (solid square), and the sin$^2$() fitting curve.

The typical phase-matching curve and conversion efficiencies of our device are presented in Fig. 2. The SFG tuning curve of the waveguide is measured by sweeping the signal wavelength from 1500 nm to 1560 nm with the pump wavelength fixed at 1950 nm. As shown in Fig. 2(a), the phase-matching signal wavelength is 1531.1 nm at room temperature at approximately 18 °C with a full width at half maximum (FWHM) of 10.5 nm. The oscillation of the curve is caused by F-P interference between the two facets, which can be eliminated by the AR coating.

The conversion efficiencies versus pumping power measured at the output port are shown in Fig. 2(b). The conversion efficiency is fitted with the model [39]

$$\eta = \eta_{max}\sin^2(\sqrt{\eta_{nor}}PL) \qquad (1)$$

where $\eta_{max}$ is the maximum efficiency, which is less than 1 due to propagation loss, P is the pump power at the output facet, $\eta_{nor}$ is the normalized conversion efficiency, and L is the quasi-phase matching length of the waveguide. The highest internal conversion efficiency is ~73% when the pump power on-chip is ~310 mW according to the fitting result, and the calculated normalized efficiency is $(2837\pm97)\%/(W\cdot cm^2)$, which is ~80% of the theoretical value of $3552\%/(W\cdot cm^2)$. Subjected to the pump power limitation, the signal depletion with a pump power of 246 mW is measured to be 14.8 dB (96.7%) by coupling the output signal into an optical spectrum analyzer and comparing the observed signal levels with the pump on and off [39].

### C. Noise Analysis

The possible noise mechanisms caused by pump laser include second harmonic generation (SHG), third harmonic generation (THG), SRS noise in fiber, SRS noise at the signal band in the waveguide, second harmonic generation-spontaneous parametric down-conversion (SHG-SPDC) noise at signal band, SHG-SRS noise at SF band, and spontaneous four-wave mixing (SFWM) noise. The SHG and THG noises can be easily removed by commercial filters, while the other noises are indistinguishable from signal or SF photons, as depicted in Fig. 3(a). The main noise source in our device is the SHG-SPDC process, especially for the high conversion efficiency condition with high power pumping. The SHG-SPDC photons that fall into the bandwidth of the QFC are upconverted to the SF band, causing noise performance deterioration.

To characterize the noise performance of the QFC chip, the SF light, pump, and nonconverted signal are separated by a longpass dichroic mirror, as shown in Fig. 3(a). A 785-nm longpass filter, a 940-nm shortpass filter, an 857-nm bandpass filter with a 30-nm bandwidth, a 0.5-nm narrowband filter centered at 857.6 nm and an etalon with a bandwidth of 0.09 nm and a free spectral range of 1.5 nm are used to clean the SF further. Then, the SF light is coupled into a multimode fiber and detected by an SPAD with a detection efficiency of ~50% and an intrinsic dark noise of ~60 cps. Taking into account the free space transmission efficiency of the noise photons, we show the total on-chip NCR dependence on the pump power in Fig. 3 (b). The inset curve with the total on-chip NCR and pump power both in the logarithmic coordinates demonstrates a slope of ~2.3 at pump power of 240 mW, showing that the total noise increases cubically with pump power in the high conversion efficiency region approximatively. This is consistent with the case mentioned above, in which the dominating noise generation process is SHG-SPDC, which has a cubic dependency on the pump power.

We further quantify the noise count from the SHG-SPDC process without other noise interference by directly importing the 975 nm light generated by a homemade 1950-nm SHG PPLN waveguide to the QFC chip. The SHG-SPDC noise count in the SF band versus the pump power is also plotted in Fig. 3(b). The proportion of SPDC noise to the total noise gradually increases as pump power gets stronger and exceeds 50% when the pump power is ~230 mW. The remaining nondominant noise can be attributed to the other noise sources, including several SRS processes and the SFWM process, which cannot be characterized separately. The SRS noises directly generated by the strong pump field in the signal band are greatly suppressed both in the fiber and waveguide because the pump laser used is largely detuned with the signal [22,32,40].

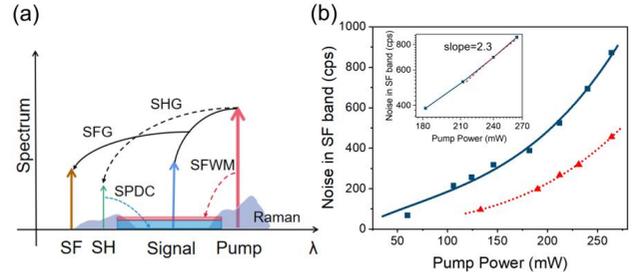

Fig. 3 (a) The main noise processes induced by a strong pump field in a QFC chip. (b) The total on-chip NCR (blue line, square) and the SHG-SPDC induced on-chip NCR (red dotted line, triangle) in the SF band over a detection bandwidth of 0.09 nm in linear coordinate. The inset shows the total on-chip NCR at high pump power in the logarithmic coordinates with the same detection bandwidth and the slope at 240 mW (red short dash).

### D. QFC of Heralded Single-Photon Source

It is generally assumed that the quantum state is preserved during the mixing process in a periodically poled LNOI nanophotonic waveguide, such as that in the conventional micron waveguide [32,41]. However, this has to be verified experimentally, since the effective refractive index and the mode profile that may affect the nonlinear process are quite different in the two platforms. Here, the nonclassical intensity correlation, one quantum property of a heralded single-photon source, is selected to verify the QFC performance [32].

We use another homemade PPLN waveguide to generate SPDC photon pairs with signal photons at the 1530-nm band and idler photons at the 1591-nm band pumped with a 780-nm CW laser, which is filtered after exiting the PPLN waveguide. The signal photons and the idler photons are separated by DWDM centered at 1530 nm, and the idler arm is filtered by a tunable filter centered at 1591 nm, as shown in Fig. 4.

First, we measured the normalized second-order cross-correlation function to describe the statistical properties of the

SPDC photon pairs in Fig. 4(a) with $g_{si}^{(2)}(\tau) = \frac{N_{s,i}}{N_s \cdot N_i \cdot \tau_b}$ [42], where $N_{s,i}$, $N_s$ ($N_i$) are the coincidence count rate and channel count rate of the signal and idler photons, and $\tau_b$ is the coincidence time window. Considering the coherence time, time jitter of the detector, and temporal resolution of the detector, we use a coincidence time window of 1500 ps and obtain $g_{si}^{(2)}(0) = 22$, showing a strong correlation well above the classical thermal statistic threshold of 2. Next, the heralded single-photon property is characterized with the idler photons played as the heralding photons, as shown in Fig. 4(b). The signal photons are split by a 50/50 beam splitter and sent to two separate channels of SNSPD. We measured the heralded self-correlation function [43]

$$g_H^{(2)}(\tau) = \frac{N_{ssi}(\tau) \cdot N_i}{N_{s1i}(0) \cdot N_{s2i}(\tau)}$$

For the special case, we are interested in where $g_H^{(2)}(\tau) \equiv g_H^{(2)}(0,\tau|0)$. $N_{s1i}(N_{s2i})$ is the double coincident count rate of the idler and signal 1 (signal 2). $N_{ssi}$ is the triple signal-signal-idler coincidence rate, and $N_i$ is the count rate of the idler arm. Fig. 5(a) shows that the lowest self-correlation value at zero delays is $g_H^{(2)}(0) = 0.14$ with a coincidence time window of 1500 ps, showing strong antibunching behavior. Finally, the 1530-nm band signal photons are upconverted to the 857-nm band by our LNOI chip, and the self-correlation function is measured with the idler photons as herald shown in Fig. 4(c). The measured self-correlation function shown in Fig. 5(b) is $g_H^{(2)}(0) = 0.23$, which is higher than that of the photon-pair source due to the noise counts introduced by the nonlinear process. However, $g_H^{(2)}(0)$ is still below the classical threshold of 0.5, proving that the single-photon nature is preserved after the frequency conversion process.

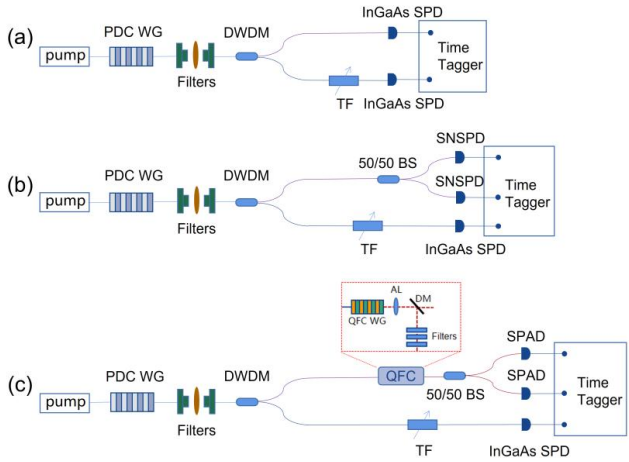

Fig. 4 The second-order correlation function measurement setups of the (a) photon pair, (b) heralded single-photon source, and (c) QFC of the herald single photon. DWDM, dense wavelength division multiplexer; BS, beam splitter; SPD, single-photon detector; TF, tunable filter.

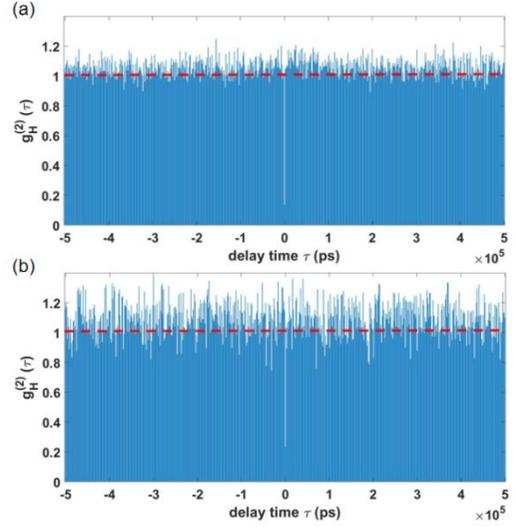

Fig. 5 Heralded self-correlation function $g_H^{(2)}(\tau)$ for (a) signal-signal and (b) SF-SF.

## 3. UPCONVERSION SPD and DISCUSSION

Here, we construct an upconversion single-photon detection based on the QFC chip. In Fig. 2(a), the signal source is replaced by a single-photon source of one million photons centered at 1531 nm consisting of two variable optical attenuators (VOAs) and a 99/1 beam splitter [33]. The filtering system and SPAD are all the same as the noise measurement setup in Fig. 2. The pump power is tuned, and the upconversion SPD system DE and NCR at different pump powers are recorded and depicted in Fig. 6. System DE is obtained by dividing the number of detected count rates after NCR subtraction by one million, which is the signal photon count rate before entering the WDM.

A system DE of 8.7% and an NCR of 300 cps are achieved at a pump power of ~240 mW on the chip, which shows preliminary practicality. The pump power can be significantly reduced by increasing the length of the chip, and consequently, the noise count would be notably depressed due to the near cubically dependent relationship. The unsatisfactory DE performance is caused by the insertion losses and efficiencies of the components, which are listed in Table 1. The major loss is a fiber-to-chip coupling loss of ~4 dB, which severely affects system performance. To reduce the coupling loss, some inverted taper designs have been proposed and experimentally realized with a coupling loss of less than 0.5 dB [44,45]. Another nonnegligible loss is the relatively low internal conversion efficiency mainly due to the transmission loss, which can be further reduced to less than 0.1 dB/cm by optimizing the fabrication process [38]. Moreover, the loss of the free space filtering module could also be further optimized. After the improvements mentioned above, a highly practical upconversion SPD with a DE of ~30% could be achieved.

In addition to the upconversion SPD, the low-noise high-efficiency on-chip frequency conversion provides possibilities for integrated QFC-based quantum systems through heterogeneous integration with other materials. Generally, a fully scalable quantum photonic chip should realize both quantum state manipulation and single-photon detection on a monolithic chip where the QFC could play crucial roles. As an important quantum state manipulation, low-noise QFC has been achieved on our LNOI chip, which could realize interfaces between long-distance quantum memories and various quantum systems. Critical but difficult, on-chip single-photon detection still needs to be explored. The integration of superconducting nanowires with LNOI has been proposed to realize single-photon detection on the telecom band [16,18], while the ultralow temperature refrigeration system may be cumbersome for some applications. Another alternative for on-chip low-noise single-photon detection is to combine an antimonide semiconductor laser as a pump laser and SPAD as an SF detector with an LNOI QFC chip by heterogeneous integration, which only requires compact temperature control.

Table. 1 Insertion losses or efficiencies of the components

| | |
|---|---|
| Insertion loss of the WDM, PC and lensed fiber | 0.5 dB |
| Fiber-to-chip coupling loss | 4.0 dB |
| Conversion efficiency | 72% |
| Filtering and space-fiber coupling loss | 1.6 dB |
| Detection efficiency of SPAD | 50% |

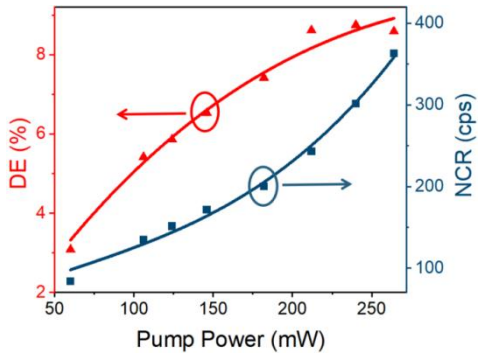

Fig. 6. DE (red line, triangle) and NCR (blue line, square) versus pump power.

## 4. CONCLUSION

In conclusion, we have achieved quantum frequency conversion in the LNOI nanophotonic platform with an internal conversion efficiency of 73% and a noise count rate of 900 cps on-chip. We verified the preservation of the quantum state after the QFC process by using the non-classical intensity correlation of a heralded single photon source as the input quantum property, which is the first verification in the LNOI platform to the best of our knowledge. Moreover, an upconversion single-photon detector in the telecom band with a system DE of 8.7% and NCR of 300 cps is developed based on the QFC chip. The on-chip low-noise and high-efficiency quantum frequency converter indicates the possibilities of chip-scale QFC-based systems in heterogeneous configurations.

The authors thank Daiying Wei for polishing the waveguide. This work is supported by National Key R&D Program of China (2018YFB0504300, 2017YFA0303902, 2017YFA0304000); National Natural Science Foundation of China (T2125010); Key R & D Plan of Shandong Province (2019JZZY010205); Natural Science Foundation of Shandong Province (ZR2019LLZ003, ZR2020LLZ007, ZR2021LLZ013); the Chinese Academy of Science; the SAICT Experts Program; the Taishan Scholar Program of Shandong Province; Quancheng Industrial Experts Program.